\documentstyle[11pt]{article}

\begin{document}
\renewcommand{\thefootnote}{\alph{footnote}}
\newcommand{\ffi}{\varphi}
\newcommand{\db}{\delta_{\rm BRST}}
\newcommand{\dd}{\delta}

\begin{titlepage}

\begin{flushright}
ITP-SB-94-21 \\
NOVEMBER 1994
arch-ive/9410116
\end{flushright}

\vspace{2cm}

\begin{center}

{\bf \Large Background gauge invariance in the antifield formalism for
theories with  open gauge algebras} \\

\vspace{2cm}

{\large Kostas Skenderis\footnote{email:kostas@max.physics.sunysb.edu} } \\

\vspace{0.5cm}

{\it Institute for Theoretical Physics, \\
State University of New York at Stony Brook,\\
Stony Brook, NY 11794-3840}

\end{center}

\vspace{2cm}

\begin{abstract}
We show that any BRST invariant quantum action with open or closed
gauge algebra has a corresponding local background gauge invariance.
If the BRST symmetry is anomalous, but the anomaly can be removed in
the antifield formalism, then the effective action possesses a local
background gauge invariance.
The presence of antifields (BRST sources) is  necessary.
As an example we analyze chiral $W_3$ gravity.
\end{abstract}

PACS: 1100, 0450.

\end{titlepage}
\pagebreak

\setcounter{footnote}{0}
\renewcommand{\thefootnote}{\arabic{footnote}}

\section{Introduction.}

Certain two dimensional field theories can be covariantly quantized
by choosing a gauge in which all gauge  fields are set equal to corresponding
background fields. An example is chiral gravity coupled to scalars
$\chi^\alpha$ with quantum Lagrangian
\begin{equation}
L_q = - \frac{1}{2} \bar{\partial} \chi^\alpha \partial \chi^\alpha
+ \frac{1}{2} h \partial \chi^\alpha \partial \chi^\alpha
+ d(h - H) - b(\bar{\partial}c - h \partial c + c \partial h). \label{gravity}
\end{equation}
The classical action is invariant
under local gauge transformations
$\delta_{g} \chi^\alpha = \epsilon \partial \chi^\alpha,
\delta_g h = \bar{\partial} \epsilon - h \partial \epsilon
+ \epsilon \partial h$,
and the quantum action is BRST invariant under
$\db \chi^\alpha = c \partial \chi^\alpha,
\db h = \bar{\partial} c - h \partial c
+ c \partial h,
\db c = - c \partial c,
\db b = d$ and $\db d= 0$.
The action in (\ref{gravity}) has then the following
background gauge invariance
\begin{eqnarray}
&\ &\delta \chi^\alpha = \epsilon \partial \chi^\alpha,
\delta h = \bar{\partial} \epsilon - h \partial \epsilon
+ \epsilon \partial h,
\delta H = \bar{\partial} \epsilon - H \partial \epsilon
+ \epsilon \partial H \nonumber \\
&\ & \delta c = \epsilon \partial c - c \partial \epsilon,
\delta b = \partial b \epsilon + 2 b \partial \epsilon,
\delta d = \partial d \epsilon + 2 d \partial \epsilon.
\end{eqnarray}
Eliminating the fields $d$ and $h$ by integrating over them in the path
integral replaces $h$ by $H$ in (\ref{gravity}), while
$\db b = T_m + T_{\rm gh}$\footnote{
$T_m = - \frac{1}{2} \partial \chi^\alpha \partial \chi^\alpha,
T_{\rm gh} = - 2 b \partial c - \partial b c$}
but the background symmetries for $\chi^\alpha, H, c,$ and $b$ do not change.

In ref. \cite{kowalski} a general class of two dimensional theories was
considered with classical gauge transformation rules
$\delta_g \phi^I = R^I_j \epsilon^j$ which were assumed to yield a closed
gauge algebra. Fixing the gauge by ${\cal L}_{\rm fix} = d_j (h^j - H^j)$,
where $h^j$ will be denoted by ``quantum gauge fields'' and $H^j$ by
``background gauge fields'', it was found that the action was background
invariant, with or without $d_j$, provided the following condition holds
\begin{equation}
\sum_{\alpha} R^i_{j,\alpha} R^\alpha_k = 0. \label{condition}
\end{equation}
Here $R^i_{j,\alpha}$ denotes $\partial R^i_j / \partial \varphi^\alpha$
with $\varphi^\alpha$ those quantum fields which are not
equal to $h^j$ (the ``matter fields''), so $\phi^I=\{h^i, \ffi^\alpha\}$.
The background transformation rules of ref. \cite{kowalski} read
\begin{eqnarray}
\dd h^i &=& R^i_j \epsilon^j, \;
\dd \ffi^\alpha = R^\alpha_j \epsilon^j, \nonumber \\
\dd C^i &=& - f^i_{\ jk} \epsilon^k C^j, \;
\dd B_i = - B_j (R^j_{l} \epsilon^l)_{,i}
\end{eqnarray}
In the example in (\ref{gravity})
$h^j$ corresponds to $h$, $\varphi^\alpha$ corresponds to $\chi^\alpha$,
and (\ref{condition}) is satisfied because $\delta_g h$ is independent
of $\chi^\alpha$.

The usefulness of local background gauge invariance of quantum actions for
Yang-Mills theories is well known. In addition, local background gauge
invariance for the type of theories considered in ref.\cite{kowalski}
allows one to compute the anomalies in each of the gauge symmetries
separately, and not only their sum as in the case of BRST anomalies\footnote{
In ref. \cite {kowalski},
the covariant action of gravity coupled to scalars had several gauge
symmetries (Einstein, Weyl, Lorentz), and requiring that each corresponding
symmetry be free from anomalies, fixed completely the ``measure'' of the
quantum fields.(``Measure'' meaning the integers $k, l, m$ in the
definition of path integral variables
$\tilde{\chi}^\alpha = e^k \chi^\alpha,
\tilde{c} = e^l c, \tilde{b} = e^m b$
where $e = {\rm det} e_\mu^m$ with  $e_\mu^m$ the vielbein field).}.
However, in all these cases the gauge algebra was assumed closed.
Although it is a widespread belief that local background gauge symmetry
exists in any BRST invariant system its precise form is not known
in general. It is the purpose of this letter to  provide
the explicit form of the local background transformations for any
anomaly-free gauge theory.

We shall use the Batalin-Vilkovisky (BV) formalism\cite{bv}.
In this language a necessary condition for local background invariance
is the existence of a gauge fermion which is invariant under the
background symmetry. We shall show that  such a gauge fixing fermion
exists for a generic gauge theory. Then the local background transformations
which leave the quantum action invariant will be constructed.
These considerations are at the level of the quantum action.
However, the BRST symmetry might be broken at the quantum level by
anomalies. Then the local background symmetry will be anomalous as well.
In some cases anomalies can be cancelled by adding suitable counterterms
to the BV action which is equivalent to adding extra terms to the quantum
action and to the transformation rules of the fields.
Then the quantum action ceases to be BRST invariant but the effective
action becomes BRST invariant. We shall show that in such cases we can also
obtain local background transformations which leave the effective action
invariant.

An example, which actually started our investigation, is
chiral $W_3$ gravity with classical Lagrangian\cite{hull}
\begin{equation}
{\cal L}_{\rm cl} =
- \frac{1}{2} \bar{\partial} \chi^\alpha \partial \chi^\alpha
+ \frac{1}{2} h \partial \chi^\alpha \partial \chi^\alpha
+ \frac{1}{3} b d^{\alpha \beta \gamma}
\partial \chi^\alpha \partial \chi^\beta \partial \chi^\gamma.
\end{equation}
It is invariant under local $\epsilon$(spin-2) and $\lambda$(spin-3)
gauge transformations, of which $h$ and $b$ are the gauge fields, respectively.
The classical gauge algebra is open: the commutator of two spin-3 gauge
transformation on the matter fields $\chi^\alpha$ is proportional to the spin-2
field equation, while the same commutator on $h$ is proportional to
(minus) the matter field equations\cite{maimi}.
Furthermore, since the classical transformation rules
of the spin-2 gauge field $h$ contains a term with the matter stress tensor
$\partial \chi^\alpha \partial \chi^\alpha$
the condition (\ref{condition}) is violated. Yet, we shall show that the
effective action for this system in the gauge $h=H$ and $b=B$ has local
background gauge invariance.

One can study either noncritical $W_3$ gravity (similar to Polyakov's
chiral gravity), or critical $W_3$ gravity. In the former case the
remaining symmetries are anomalous and one should not introduce ghosts,
while in the latter case all anomalies cancel and one can then use the
Batalin-Vilkovisky (BV) formalism\footnote{
Let us mention, however, recent attempts to quantize anomalous theories
within the framework of BV\cite{anomalies}.}.
To cancel all spin-3 anomalies,
one should use a matter system with central charge $c=100$, but
one cannot take 50 scalars doublets to achieve
this since the $W_3$ algebra is nonlinear\cite{hull,tensor}.
Rather, as first proposed in
\cite{maimi} and achieved in \cite{pope}, one should add
``background charges''\cite{romans} to the action, i.e., terms of the form
$\sqrt{\hbar} h \partial^2 \ffi + \hbar b \partial^3 \ffi + \cdots$.
The quantum action is then no longer BRST invariant at order $\hbar$,
but the effective action becomes BRST invariant to all orders in $\hbar$.
We shall  show that when the antifields are present the effective action
possesses background gauge invariance on top of the rigid BRST symmetry.

In section 2, we prove that any anomaly-free theory with open or closed
gauge algebra possesses background gauge invariance. Then, in section 3,
we analyze the critical $W_3$ gravity as an example of the general
formalism. Two appendices follow. In appendix A we present details
of the BV quantization of $W_3$ gravity with background charges.
In particular, we explicitly show how to obtain the terms depending on
``background charges'' from the BV approach.
Finally, appendix B contains the (rather lengthy) background
transformation rules which leave the $W_3$ system invariant.

\section{The general case.}

Let us now drop the examples and consider the general case. At the end
we shall specialize the results to background transformations for the
chiral $W_3$ system. We consider theories with an open gauge algebra,
so it is natural to work with the Batalin-Vilkovisky (BV) formalism.
In the BV formalism one introduces
for each field $\phi^A$ (in our case,
the fields $h^i$ which will be gauge fixed by setting them equal to
a background value, the corresponding ghosts $C^i$
and the rest of the fields $\ffi^\alpha$)
an antifield\footnote{
Antifields is another word for sources for BRST variations. They were
first introduced by Zinn-Justin\cite{zinn}
 in order to control the renormalization
of the (non-linear) BRST variations and used in the study of renormalization
of Yang-Mills theories.
}
$\phi^*_A$ which has opposite statistics from the field $\phi^A$.
Fields and antifields are  conjugate
variables in the sense that they obey the canonical relations
\begin{equation}
(\phi^A, \phi^*_B) = \delta^A_{\ B}; \;
(\phi^A, \phi^B) = (\phi^*_A, \phi^*_B) = 0,
\end{equation}
where $(\ ,\ )$ is the BV antibracket defined by
\begin{equation}
(A, B) = \partial A/ \partial \phi^A \frac{\partial}{\partial \phi^*_A}B
- \partial A/ \partial \phi^*_A \frac{\partial}{\partial \phi^A}B.
\end{equation}
$\frac{\partial}{\partial \phi}$ and $\partial\  /\partial \phi$ denote
left and right derivative, respectively.
A ghost number is assigned to each field and antifield. The ghost number
of the fields $h^i$ and $\varphi^\alpha$ is equal to zero, of the ghosts $C^i$
it is equal to one\footnote{
We consider only irreducible theories.
Our results can be straightforwardly extended to the
case of reducible theories.},
of the antighosts $B_i$ it is equal to minus one
and for the antifields we have $gh(\Phi^*_A) = - gh(\Phi_A) - 1$.
The BV action, which is a bosonic functional of ghost number zero,
is given as a power series in antifields\footnote{
Alternatively, one can expand the action in
antighost (sometimes called also antifield) number\cite{henneaux,vandoren}.
The antighost number is equal to zero for fields and to minus the ghost
number for antifields.
This formulation provides a convenient framework for the study of
the antibracket cohomology\cite{henneaux2}.
However, we will stick to the original version
of the method since it is conceptually simpler and
for practical calculations there is no real difference between the
two formulations.}
$S = S^0 + S^1 + S^2 + \cdots$, where $S^i$ contains $i$ antifields,
and it is determined by solving the ``master equation''
\begin{equation}
(S, S) = 0. \label{master}
\end{equation}
The gauge fixing is performed at the very end by adding to the action
the so-called non-minimal term $d_i B^{*i}$,
\begin{equation}
S_{\rm gf} = S + d_i B^{*i},
\end{equation}
where $d_i$ are the BRST auxiliary fields and $B^{*i}$ is the
antifield antighost.
Then a canonical transformation is performed
\begin{eqnarray}
(\phi^A)' &=& e^{\psi} \phi^A \equiv \phi^A + (\psi,\phi^A) +
\frac{1}{2} (\psi, (\psi,\phi^A)) +  \cdots, \label{gff} \\
(\phi^*_A)' &=& e^{\psi} \phi^*_A \equiv \phi^*_A + (\psi,\phi^*_A) +
\frac{1}{2} (\psi, (\psi,\phi^*_A)) +\cdots, \label{gfa}
\end{eqnarray}
where $\psi$ is the gauge fixing fermion.
One may check that after the gauge fixing
$S_{\rm gf} (\phi^A, \phi^*_A = 0)$ gives a well-defined theory
(propagators are well-defined for all fields). However, we
choose to keep the antifields (BRST sources) present.
In this way the BRST transformations
are always nilpotent independently of whether the gauge algebra is closed
or open. These transformation rules read
\begin{eqnarray}
\db \phi^A &=& (\phi^A, S_{\rm gf})
= \frac{ \partial S_{\rm gf}}{\partial \phi_A^*}, \label{brstf}\\
\db \phi_A^* &=& (\phi_A^*, S_{\rm gf})
= - \frac{\partial S_{\rm gf}}{ \partial \phi^A}. \label{brsta}
\end{eqnarray}

We wish to examine whether the BRST invariant system is also invariant under
local background gauge transformations. In the case of ordinary
Yang-Mills theories in four dimensions
these transformations are local gauge transformations for the background
gauge fields and  homogeneous transformations for the rest of the
fields\cite{abbott}. The standard way to introduce the background
fields is to split the gauge fields $h^i$ into a background part $H^i$
and a quantum part $h_q^i$ (we are suppressing the spacetime index),
\begin{equation}
h^i = h_q^i + H^i. \label{splitting}
\end{equation}
The gauge transformations are also split into quantum gauge transformations,
which will be gauge fixed, and background gauge transformations,
(the latter are from now on always denoted by $\dd$)
\begin{eqnarray}
&\ & \delta_q h^i_q = D(h_q+H) \eta^i; \hspace{0.5cm}  \delta_q H^i = 0, \\
&\ & \delta h^i_q = f^i_{\ jk} h^k_q \eta^j; \hspace{0.5cm}
\delta H^i = D(H) \eta^i,
\end{eqnarray}
where $D(h)\eta^i=\partial \eta^i + f^i_{\ jk} h^k \eta^j$ is the covariant
derivative of  the gauge field $h^i$.
Then a gauge fixing condition is chosen which  transforms covariantly
under the background gauge transformations. The most common choice is
$D(H)h_q^i=0$. This insures that the gauge fixing and the Faddeev-Popov part
of the action are invariant under the background gauge transformations.
If one then changes variables from $h^i_q$ to $h^i$ in the path integral,
the classical action only depends on $h^i$ and the background fields
enter only through the gauge fixing condition.
In the BV language this means that only the gauge fermion
depends on the background field. In the Yang-Mills case we achieved
background invariance by choosing a gauge fixing condition that transforms
covariantly. This translates in the BV language into a gauge fermion
which is invariant under background transformations. It is precisely
this point which will be our guiding principle in the search for
background invariance in theories with open gauge algebras.

We restrict our attention to gauge fermions of the form
\begin{equation}
\psi = B_i F^i, \label{fermion}
\end{equation}
where $B_i$ is the antighost and $F^i$ is the gauge fixing condition.
This choice of the gauge fermion leads to a delta function type gauge
fixing. If we would add an extra term of the form $B_i d_i$ in the
right hand side of (\ref{fermion}) we would get a Gaussian type gauge
fixing. Since we do not expect that in a generic theory we can find
such $F^i$ that transforms covariantly under the background symmetry
(so that the $F^2$ term in the action will be invariant) we will not
consider this case. On the other hand, the choice (\ref{fermion})
allows for a generic solution. Indeed, if we declare that the background
gauge fields transform the same way as the gauge fields but with the gauge
fields in the transformation rules  replaced by background fields then
their difference transforms homogeneously,
\begin{equation}
\dd (h^i - H^i) = f^i_{\ jk} (h^k - H^k) \eta^j,
\end{equation}
where $f^i_{\ jk}$ is, in general, field dependent and with no symmetry in
the indices $i, j, k$. It is clear that the choice\footnote{
A little bit more general choice is also possible, namely
$F^i = a^i_{\ j}(h^i - H^i)$, where $a^i_{\ j}$ is either constant invertible
matrix or differential operator. In the first case appropriate factors
of  $a^i_{\ j}$ and its inverse should be added to the right hand side of
(\ref{Brule}), whereas in the second one, we integrate by part $a^i_{\ j}$
so that it acts on $B_i$ (schematically, $B_i \rightarrow a^{\ j}_i B_j$
after the integration by parts).
Then $B_i$ should be replaced by $a^{\ j}_i B_j$ in (\ref{Brule}).}
\begin{equation}
F^i = (h^i - H^i),
\end{equation}
leads to a background invariant gauge fermion if we choose  the
transformation rules of the antighost field appropriately, namely
\begin{equation}
\dd B_i = - B_j f^j_{\ ki} \eta^k (-1)^{\epsilon(h) \epsilon(\eta)},
\label{Brule}
\end{equation}
where $\epsilon(A)$ is equal to 0 (1) if $A$ is a boson (fermion).

Next  we perform the canonical transformation.
{}From (\ref{fermion}), (\ref{gff}) and (\ref{gfa}) we find
\begin{eqnarray}
(B^{*i})' &=& B^{*i} + h^i - H^i, \\
(h^*_i)'  &=& h^*_i + B_i.
\end{eqnarray}
Since the antighost $B_i$ always appears together with the antifield
$h^*_i$, for notational simplicity we will denote their sum by  $\hat{h}^*_i$,
\begin{equation}
\hat{h}^*_i = h^*_i + B_i. \label{id1}
\end{equation}
(Note, however, that the antighost $B_i$ is a field which we integrate over
in the path integral whereas $h^*_i$ is an external field).
The field $\hat{h}^*_i$  transforms under BRST (use (\ref{brstf}) and
(\ref{brsta})) as follows
\begin{equation}
\db \hat{h}^*_i = \db h^*_i + \db B_i = (h^*_i, S)\Big|_{h^*_i=\hat{h}^*_i}.
\end{equation}
Similarly, we denote the difference between the antifield antighosts $B^{*i}$
and the background fields $H^i$ by $\hat{H}^i$,
\begin{equation}
\hat{H}^i = H^i - B^{*i}. \label{id2}
\end{equation}
Furthermore, let us denote by $\hat{S}$ the solution of the
master equation (\ref{master})
with the antifields $h_i^*$ replaced by $\hat{h}^*_i$.
In this notation the gauge fixed action reads
\begin{equation}
S_{\rm gf} = \hat{S} + d_i (h^i - \hat{H}^i). \label{gfaction}
\end{equation}

We now claim that the BRST invariant action in (\ref{gfaction}) is
invariant under the following transformations
\begin{eqnarray}
\dd \phi^A &=&
\partial (\phi^A, \hat{S})/\partial C^i \eta^i,
\label{backfield} \\
\dd \phi^*_A &=&
\partial (\phi^*_A, \hat{S})/\partial C^i \eta^i,
\label{backanti} \\
\dd \hat{H}^i &=& \dd h^i \Big|_{h^i = \hat{H}^i}, \\
\dd d_i &=& - d_k f^k_{\ ji} \eta^j (-1)^{\epsilon(h) \epsilon(\eta)},
\label{backd}
\end{eqnarray}
where $\phi^A=\{h^i, \varphi^\alpha, C^i\}$ and
$\phi^*_A=\{\hat{h}^*_i, \varphi^*_\alpha, C^*_i\}$ and
$\eta^i(x)$ is a local parameter.
Notice that we have already determined the transformation rules of the
antighost in equation (\ref{Brule}). Hence (\ref{backanti}) uniquely
determines the transformation rules of the antifield $h^*_i$.

The non-minimal term
\begin{equation}
d_i (h^i - \hat{H}^i),
\end{equation}
is invariant by itself
whereas for the minimal terms we have
\begin{eqnarray}
\dd \hat{S} &=& \partial \hat{S}/\partial \phi^A \dd \phi^A
+ \partial \hat{S}/\partial \phi^*_A \dd \phi^*_A \nonumber \\
&=& \frac{1}{2}  \partial (\hat{S}, \hat{S}) /\partial C^i \eta^i = 0,
\label{proof}
\end{eqnarray}
since $(\hat{S}, \hat{S})=0$ by construction.

Notice that unless the gauge algebra is closed with field independent
structure constants the antifields cannot be set consistently equal to
zero (i.e. $\dd \phi^*_A \neq 0$ when $\phi^*_A=0$). In contrast,
one can project the BRST transformations rules
for the fields (\ref{brstf}) onto  hypersurface $\phi^*_A=0$
and set the BRST variation of the antifields consistently equal to zero
(in fact this is precisely what is called
``BRST variation'' of fields and antifields
in the original papers of Batalin and Vilkovisky\cite{bv}).

This concludes our treatment of theories without anomalies. We now extend
our consideration to theories with anomalies.
If a theory is anomalous, then
\begin{equation}
(\Gamma, \Gamma) = i \Delta \cdot \Gamma, \label{anomaly}
\end{equation}
where $\Gamma$ is the effective action
\begin{equation}
\Gamma = S + \hbar \Gamma_1 + \hbar^2 \Gamma_2 + \cdots,
\end{equation}
and $\Gamma \cdot \Delta$ denotes the set of all 1PI graphs with an insertion
of the composite operator $\Delta$,
\begin{equation}
\Delta = \hbar \Delta_1 + \hbar^2 \Delta_2 + \cdots.
\end{equation}
Clearly, the breakdown of BRST symmetry implies
the breakdown of the background symmetry as well.
The anomaly in the background symmetry is then given by
\begin{equation}
\Delta_{\rm back}
= \frac{1}{2} \partial \Delta /\partial C^i \eta^i.
\end{equation}

In certain two dimensional theories (e.g. $W_3$-gravity),
one can make the effective action BRST invariant by adding
background charges to the action. The quantum action becomes then a series
in half-integer powers of $\hbar$
\begin{equation}
S = S_0 + \hbar^{1/2} S_{1/2} + \hbar S_1 + \cdots, \label{backaction}
\end{equation}
where $S_0 = S^0 + S^1 + \cdots$.
The new terms serve as counterterms which cancel the anomalies.
The effective action becomes also a power series in half-integer
powers of $\hbar$.
The Zinn-Justin equation, $(\Gamma, \Gamma)=0$, reads
\begin{eqnarray}
(S_0, S_0) &=& 0, \label{oldmaster} \\
(S_0, S_{1/2}) &=& 0, \label{1/2level} \\
(S_{1/2}, S_{1/2}) + 2 (S_0, S_1) &=& i \Delta_1,
\hspace{0.5cm} \mbox{etc.,} \label{1level}
\end{eqnarray}
where we have assumed that the anomaly appears for first time at order $\hbar$.
Equation (\ref{1level}) involves only
local quantities because the anomaly is always local when it first appears.
At higher order in $\hbar$ the Zinn-Justin equation involves non-local
expressions and loop calculations.

Equation (\ref{oldmaster}) is just the old master equation (\ref{master}).
The new equations are solved recursively in the number of
antifields (see appendix A).
Equations (\ref{oldmaster}-\ref{1level})
together with the Wess-Zumino consistency condition,
\begin{equation}
(\Gamma, \Delta \cdot \Gamma) = 0, \label{W-Z}
\end{equation}
determine the anomaly-free quantum action (if it exists)
up to BRST exact terms once the antifield independent part of the classical
action in (\ref{backaction}) and  the anomaly in (\ref{anomaly}) is
given\cite{vandoren}.

It is clear from (\ref{1level}) that the classical action is not BRST
invariant any more.
However, the effective action is BRST invariant to all orders
\begin{equation}
(\Gamma, \Gamma) = 0.
\end{equation}
This last relation ensures also the local background invariance of the
effective action since
\begin{equation}
\dd \Gamma = \frac{1}{2}  \partial (\Gamma, \Gamma)/ \partial C^i \eta^i
=0.
\end{equation}
The proof is identical to that in (\ref{proof}).

\section{Example: chiral $W_3$ gravity}

We now apply the general formalism to chiral $W_3$ gravity.
The classical action is given by\cite{hull}
\begin{equation}
I = \frac{1}{\pi}
\int d^2x [- \frac{1}{2} \bar{\partial} \chi^\alpha \partial \chi^\alpha
+ \frac{1}{2} h \partial \chi^\alpha \partial \chi^\alpha
+ \frac{1}{3} b d^{\alpha \beta \gamma}
\partial \chi^\alpha \partial \chi^\beta \partial \chi^\gamma], \label{iaction}
\end{equation}
where $\chi^\alpha$ ($\alpha=1,\ldots, n$) are scalar fields, and
$h$ and $b$ are spin-2 and spin-3 gauge fields, respectively.
This action is invariant under the following local gauge transformations
\begin{eqnarray}
\dd_g \chi^\alpha &\equiv& R^\chi_j \eta^j =
\epsilon \partial \chi^\alpha
+ \lambda \partial \chi^\beta \partial \chi^\gamma d^{\alpha \beta \gamma}, \\
\dd_g h &\equiv& R^h_j \eta^j =
\bar{\partial} \epsilon - h \partial \epsilon
+ \epsilon \partial h + (\lambda \partial b - b \partial \lambda)
 \partial \chi^\alpha \partial \chi^\alpha, \label{hgauge}\\
\dd_g b &\equiv& R^b_j \eta^j =
\epsilon \partial b - 2 b \partial \epsilon
+ \bar{\partial} \lambda - h \partial \lambda + 2 \lambda \partial h,
\label{bgauge}
\end{eqnarray}
where $\eta^j=\{\epsilon, \lambda\}$,
provided that the $d^{\alpha \beta \gamma}$ symbol is
completely symmetric and satisfies the following identity\cite{hull}
\begin{equation}
d^{\alpha (\beta \gamma} d^{\delta) \epsilon \alpha }
= \dd^{(\beta \gamma} \dd^{\delta) \epsilon}.
\end{equation}

It is well known that the gauge algebra of $W_3$ gravity is open.
Moreover, the condition (\ref{condition}) is not satisfied.
Indeed,
\begin{equation}
(R^h_{\lambda} \lambda)_{,\chi} R^\chi_j \eta^j =
(\lambda \partial b - b \partial \lambda) \partial \chi^\alpha
\partial(\epsilon \partial \chi^\alpha
+ \lambda \partial \chi^\beta \partial \chi^\gamma d^{\alpha \beta \gamma})
\neq 0.
\end{equation}
Hence, the methods of ref.\cite{kowalski} are not applicable. In order
to apply the formalism of the previous section we need as a starting
point the BV quantized theory.

The BV quantization of $W_3$ without background charges was carried out
in ref.\cite{maimi}. There it was also shown that anomalies cannot be removed
by adding to the action counterterms of order $\hbar$ and it was suggested
that introduction of background charges might lead to anomaly-free theory.
The anomaly-free $W_3$ gravity was first constructed in ref.\cite{pope}.
The extension of the BV scheme to include the case of background charges
was given in ref.\cite{vandoren}. However, there  a modified version
of the BV scheme was used (see footnote 6).
In the appendix we present a  derivation of
the same results based on the original BV scheme.
Here we just quote the final answer for the quantum action.
\begin{eqnarray}
S &=& \frac{1}{\pi}
\int d^2x \Big[ {\cal L}_{\rm cl}
+\sqrt{\hbar} [h \alpha_\alpha \partial^2 \chi^\alpha
+ e_{\alpha \beta} b \partial \chi^\alpha \partial^2 \chi^\beta] \nonumber \\
&\ & \hspace{1.1cm}
+\hbar f_\alpha b \partial^3 \chi^\alpha \nonumber \\
&\ &+ h^* \big[ \bar{\partial} c + c \partial h  - h \partial c
+ (\gamma \partial b - b \partial \gamma )
\partial \chi^\alpha \partial \chi^\alpha \nonumber \\
&\ & \hspace{0.8cm}
+\sqrt{\hbar} [2 (\gamma \partial b - b \partial \gamma)
\alpha_\alpha \partial^2 \chi^\alpha -
2 \alpha_\alpha \partial \chi_\alpha^* \gamma \partial \gamma] \nonumber \\
&\ & \hspace{0.8cm}
+\hbar \frac{25}{24}
(2 \gamma \partial^3 b - 3 \partial \gamma \partial^2 b
+3 \partial b \partial^2 \gamma - 2 b \partial^3 \gamma) \big] \nonumber \\
&\ &+ b^*[c \partial b - 2 b \partial c + \bar{\partial} \gamma
+2 \gamma \partial h - h \partial \gamma] \nonumber \\
&\ &+ c^*[-c \partial c
- \gamma \partial \gamma \partial\chi^\alpha \partial \chi^\alpha]
 \nonumber \\
&\ & \hspace{0.8cm}
-\sqrt{\hbar} 2 \alpha_\alpha \gamma \partial \gamma \partial^2 \chi^\alpha
-\hbar \frac{25}{24}
(2 \gamma \partial^3 \gamma - 3 \partial \gamma \partial^2 \gamma)
\nonumber \\
&\ &+ \gamma^*[-c\partial\gamma + 2 \partial c \gamma] \nonumber \\
&\ &+ \chi^*_\alpha \big[c \partial \chi^\alpha
+ \gamma \partial \chi^\beta \partial  \chi^\gamma d^{\alpha \beta \gamma}
+2 h^* \gamma \partial \gamma \partial \chi^\alpha \nonumber \\
&\ & \hspace{0.8cm}
+\sqrt{\hbar} [-\alpha_\alpha \partial c +
(e_{\alpha \beta} - e_{\beta \alpha}) \gamma \partial^2 \chi^\beta
-e_{\beta \alpha} \partial \gamma \partial \chi^\beta] \nonumber \\
&\ & \hspace{0.8cm}
+\hbar f_\alpha \partial^2 \gamma \big] \Big], \label{w3action}
\end{eqnarray}
where  $c$ and $\gamma$ are the ghosts for the $\epsilon$ and $\lambda$
symmetry, respectively, $ h^*, b^*$,
\linebreak $\chi^*_\alpha, c^*,
\gamma^*$ the  antifields, and
$\alpha_\alpha, e_{\alpha \beta}$ and $f_\alpha$ constants which satisfy
the following complete set of equations\cite{romans}
\begin{eqnarray}
&\ &e_{(\alpha \beta)} - d^{\alpha \beta \gamma} \alpha_\gamma =0,
\label{r1} \\
&\ &d^{\alpha \beta \delta}(e_{\delta \gamma} -  e_{\gamma \delta})
+ 2 d^{\gamma \delta  (\alpha} e_{\beta) \delta} =
2 \delta_{\alpha \beta} \alpha_\gamma, \label{r2} \\
&\ &3 f_\alpha - \alpha_\beta e_{\beta \alpha} = 0 \label{r3} \\
&\ &d^{\alpha \beta \beta} - 6 e_{\alpha \beta} \alpha_\beta
+ 6 f_\alpha = 0 \label{r4} \\
&\ &d^{\alpha \gamma \delta} d^{\beta \gamma \delta}
+ 6 d^{\alpha \beta \gamma} f_\gamma - 3 e_{\alpha \gamma} e_{\beta \gamma}
= \frac{261}{8} \delta_{\alpha \beta} \label{r5} \\
&\ &12 \alpha_\alpha \alpha_\alpha - n + 100 = 0. \label{r6}
\end{eqnarray}
Equations (\ref{r1}-\ref{r5}) were originally derived
by requiring that the $W_3$ algebra
closes at the quantum level in the presence of background
charges\cite{romans}.
Then additional effort was necessary
in order to obtain an action which possesses
the $W_3$ symmetry at the quantum level\cite{pope}.
However, this method did not provide the
transformation rules which leave the action invariant.
In contrast, the BV scheme provides the action, the transformation rules
of fields and antifields and the above set of equations at the same time.
(Of course, one can also use the Hamiltonian formalism of
Batalin-Fradkin-Vilkovisky (BFV) to obtain both the action and the
transformation rules. For a recent work in this direction see
ref.\cite{stelle}\footnote{
The results of ref.\cite{stelle} follow directly from Hamiltonian
BFV formalism. I am indebted to Peter van Nieuwenhuizen and to
Andrew Waldron for pointing out this to me.}.)

We gauge fix by adding to the action the non-minimal term
\begin{equation}
d_{h} B^{*h} + d_{b} B^{*b},
\end{equation}
and then performing a canonical transformation with gauge fermion,
\begin{equation}
\psi = B_{h} (h - H) + B_{b} (b - B),
\end{equation}
where $H$, $B$ are background fields,
$B_{h}, B_{b}$ the antighosts for the spin-2 and spin-3
symmetry, respectively, $B^{*h}$, $B^{*b}$ the corresponding
antifields and $d_{h}, d_{b}$ the BRST auxiliary fields.
After the gauge fixing the action becomes
\begin{equation}
S_{\rm gf} = S(h^*=\hat{h}^*, b^*=\hat{b}^*)
+ d_h (h - \hat{H}) + d_b (b - \hat{B}, \label{fixed})
\end{equation}
where the hatted ``antifields'' have been defined in (\ref{id1}), and
the hatted background fields in (\ref{id2}).

Once we have the BRST invariant theory, the equations (\ref{backfield}-
\ref{backd}) give the background transformation rules under which the
effective action is invariant. The explicit form of these (lengthy)
transformation rules is given in appendix B. Here we just mention that
the presence of antifields is necessary in order the system to have
background invariance.
Indeed, since $\hat{h}^* = h^* + B_h$, the variation
of $\hat{h}^*$ (\ref{hhat}) is never zero (similarly $\dd \hat{b}^* \neq 0$).
Hence, from (\ref{chi}) and (\ref{gamma}) we infer that
$\dd \chi^*_\alpha \neq 0$ and $\dd \gamma^* \neq 0$.
And from that it follows that $\dd c^* \neq 0$ as well.
Therefore, only the antifields $h^*$ and $b^*$ can be consistently set
equal to zero.

Having established the background symmetry we can write down Ward identities.
Moreover, since BRST anomalies imply anomalies in the background symmetry,
one may choose to study the background symmetry instead of the BRST one
(see, for example ref.\cite{hull2}).
However, this has the serious disadvantage that antifields must
necessarily be present. In contrast, in the case of the BRST
anomalies one has to calculate only the antifield independent part
of the anomaly since the rest follows from the Wess-Zumino consistency
condition\cite{vandoren}.
\newline
\\
{\bf Acknowledgement:} I would like to thank Peter van Nieuwenhuizen
for suggesting me this problem, for carefully reading the manuscript
and for discussions. Discussions  with J. de Boer, M. Ro\v{c}ek and
K. Stelle are also acknowledged.

\appendix

\section{Appendix}

\setcounter{equation}{0}

\renewcommand{\theequation}{\Alph{section}.\arabic{equation}}

Equation (\ref{1/2level}) is solved by writing $S_{1/2}$ as a polynomial
in antifields,
\begin{equation}
S_{1/2} = S_{1/2}^0 + S_{1/2}^1 + S_{1/2}^2 + \cdots.
\end{equation}
Then (\ref{1/2level}) becomes the system of equations
\begin{eqnarray}
(I, S_{1/2}^1) + (S_{1/2}^0, S_0^1) & = & 0,  \label{1/2-1} \\
(S_0^1, S_{1/2}^1) + (I, S_{1/2}^2) + (S_0^2, S_{1/2}^0) & = & 0,
\label{1/2-2} \hspace{0.5cm} \mbox{etc.,}
\end{eqnarray}
where $I$ is the classical gauge invariant action, for example (\ref{iaction}).
This system determines the antifield dependent part of $S_{1/2}$ once
the antifield independent part, $S_{1/2}^0$, is given.
Equation (\ref{1/2-1}) provides also a consistency check. Clearly,
$(I, S_{1/2}^1)$ is proportional to field equations. So, in order the
system to have solution $(S_{1/2}^0, S^1_0)$ has to be proportional
to field equations.

For the case of the $W_3$ gravity the antifield independent part of
$S_{1/2}$ is taken to be
\begin{equation}
S_{1/2}^0 = \frac{1}{\pi} (h \alpha_\alpha \partial^2 \chi^\alpha
+ e_{\alpha \beta} b \partial \chi^\alpha \partial^2 \chi^\beta).
\label{counter1/2}
\end{equation}
To motivative this choice note that the above expression is nothing but
$h T + b W$, where $T$ and $W$ are the spin-2 and spin-3 currents, with
one matter field removed from each current and then the indices
contracted appropriately with the help of new constants.
This justify also the order in $\hbar$ of that term.

Equation (\ref{1/2-1}) then requires that (\ref{r1}) and (\ref{r2})
hold and determines $S_{1/2}^1$ up to terms proportional to
ghost antifields. These terms are determined at the next level
together with the $S^2_{1/2}$. The results are
\begin{eqnarray}
S_{1/2}^1 &=& \frac{1}{\pi} [2 h^* (\gamma \partial b - b \partial \gamma)
\alpha_\alpha \partial^2 \chi^\alpha \nonumber \\
&\ &+\chi^*_\alpha [-\alpha_\alpha \partial c +
(e_{\alpha \beta} - e_{\beta \alpha}) \gamma \partial^2 \chi^\beta
-e_{\beta \alpha} \partial \gamma \partial \chi^\beta] \nonumber \\
&\ &-2 c^* \alpha_\alpha \gamma \partial \gamma \partial^2 \chi^\alpha],\\
S_{1/2}^2 &=&
- \frac{1}{\pi}
2 \alpha_\alpha h^* \partial \chi_\alpha^* \gamma \partial \gamma.
\end{eqnarray}
The higher order equations (in antifields) are satisfied identically.

At the next level the input is the antifield independent parts
of $S_1$ and $\Delta_1$. Equation (\ref{1level}) becomes
\begin{eqnarray}
&\ &2[(I, S_1^1) + (S_0^1, S_1^0) + (S_{1/2}^0, S_{1/2}^1)] =i \Delta_1^0
 \label{1-1} \\
&\ &2[(I, S_1^2) + (S_0^1, S_1^1) + (S_0^2, S_1^0) + (S_{1/2}^0, S_{1/2}^2)]
+ (S_{1/2}^1, S_{1/2}^1) = \nonumber \\
&\ & \hspace{7cm} = i \Delta_1^1,
\hspace{0.5cm} \mbox{etc.} \label{1-2}
\end{eqnarray}
Equation (\ref{1-1}) determines $S_1^1$ (and similarly with (\ref{1/2-1})
provides a consistency condition), but equation (\ref{1-2}) seems that
it has two unknowns, namely $S_1^2$ and $\Delta_1^1$. However, $\Delta_1^1$
can be determined from the Wess-Zumino consistency condition (\ref{W-Z}).
At the one-loop level and up to two-antifields  (\ref{W-Z})  reads
\begin{eqnarray}
&\ &(I, \Delta_1^1) + (S_0^1, \Delta_1^0) = 0,  \label{wz1} \\
&\ &(I, \Delta_1^2) + (S_0^1, \Delta_1^1) + (S_0^2, \Delta_1^0) = 0.
\label{wz2}
\end{eqnarray}

One continues in a similar fashion going up in $\hbar$ with step $\hbar^{1/2}$.
Of course, in order to determine the higher order terms in the effective
action one needs to perform loop calculations. For example, $\Gamma_{3/2}$
is determined by one-loop calculations with vertices from $S_{1/2}$.
If a new anomaly appear at this level it will be local. Then one looks
for a possible counterterm $S_{3/2}$ such that the new anomaly is cancelled.
At each level in $\hbar$ the resulting equations are solved perturbatively
in antifields. This procedure terminates when all equations are satisfied
identically.

Going back to $W_3$ we need as input the antifield independent part
of the one-loop anomaly and the  antifield independent part of $S_1$.
For the latter we take
\begin{equation}
S_1^0 = \frac{1}{\pi} f_\alpha b \partial^3 \chi^\alpha.
\end{equation}
The motivation is the same as for the case of $S_{1/2}^0$ (this time we
remove two scalar fields).
The antifield independent one-loop anomaly can be obtained by either
calculating Feynman graphs\cite{maimi}
or regularizing the path integral\cite{vandoren}.
The result reads
\begin{eqnarray}
i \Delta_1^0 &=& \frac{n-100}{6\pi} c \partial^3 h
- \frac{1}{3\pi} d^{\alpha \alpha \beta} \partial \chi^\beta
(b \partial^3 c - \gamma \partial^3 h) \nonumber \\
&+&\frac{2}{3\pi}
d^{\alpha \gamma \delta} d^{\beta \gamma \delta}
\gamma \partial \chi^\alpha \partial^3 (b \partial \chi^\beta)
-\frac{2}{\pi} (\gamma \partial b - b \partial \gamma)
\partial^3 \chi^\alpha \partial\chi^\alpha \nonumber \\
&+&\frac{1}{3\pi}
\partial \chi^\alpha \partial\chi^\alpha
(-5 \gamma \partial^3 b + 5 \partial^3 \gamma b
+12 \partial \gamma \partial^2 b - 12 \partial^2 \gamma \partial b).
\end{eqnarray}
With these data we solve (\ref{1-1}). This requires that equations
(\ref{r3}-\ref{r6}) hold\footnote{
In fact, one gets as necessary conditions the equations
(\ref{r3}, \ref{r4}, \ref{r6}) and
$ e_{\alpha \beta} e_{\beta \alpha} = y \delta_{\alpha \beta}$, \newline
$d^{\alpha \gamma \delta} d^{\beta \gamma \delta}
+ 6 d^{\alpha \beta \gamma} f_\gamma - 3 e_{\alpha \gamma} e_{\beta \gamma}
= 3z \delta_{\alpha \beta}$,
and
$6 d^{\alpha \beta \gamma} f_\gamma + e_{\gamma \alpha} e_{\gamma \beta}
- 4 \alpha_\alpha \alpha_\beta = (3z+4y-2) \delta_{\alpha \beta}$.
Then self-consistency of the system determines
$z=87/8$, $y=-49/8$.}
and yields $S^1_1$ up to ghost antifields terms,
\begin{eqnarray}
S_1^1 &=& h^* \frac{25}{24\pi}
(2 \gamma \partial^3 b - 3 \partial \gamma \partial^2 b
+3 \partial b \partial^2 \gamma - 2 b \partial^3 \gamma)  \nonumber \\
&+&\frac{1}{\pi} \chi_\alpha^* f_\alpha \partial^2 \gamma
+ \cdots, \label{s11}
\end{eqnarray}
where the dots indicate the ghost antifields terms.
These are determined at the next level (equation (\ref{1-2})).
To solve equation (\ref{1-2}) we need $\Delta_1^1$. This is obtained
by solving equation (\ref{wz1}) and (\ref{wz2}) (to get the $c^*$ part).
\begin{eqnarray}
i \Delta_1^1 &=& \frac{1}{3\pi} c^* \gamma \partial \gamma
(n \partial^3 c + 2 d^{\alpha \alpha \beta}
\partial^3 (\gamma \partial \chi^\beta)) \nonumber \\
&+& \frac{1}{3\pi} h^* \big[
(\gamma \partial b - b \partial \gamma)
(-n \partial^3 c - 2 d^{\alpha \alpha \beta}
\partial^3 (\gamma \partial \chi^\beta)
\nonumber \\
&\ & \hspace{0.5cm} +\gamma \partial \gamma
(n \partial^3 h +
2 d^{\alpha \alpha \beta} \partial^3 (b \partial \chi^\beta))\big]
\nonumber \\
&+& \frac{1}{3\pi} \chi_\alpha^* \big[
- d^{\alpha \beta \beta} \gamma \partial^3 c
-2 d^{\alpha \gamma \delta} d^{\beta \gamma \delta}
\gamma \partial^3 (\gamma \chi^\beta)
\nonumber \\
&\ & \hspace{0.5cm}
+12 \partial (\gamma \partial \gamma \partial^2 \chi^\alpha)
+18 \partial^2 \gamma \partial \gamma \partial \chi^\alpha
+ 16 \gamma \partial^3 \gamma \partial \chi^\alpha \big],
\end{eqnarray}
which agrees with the result given in ref.\cite{vandoren}.
Now, equation (\ref{1-2}) yields the undetermined part from the level 1,
and vanishing result for $S_1^2$,
\begin{equation}
S_1^1 = \cdots - \frac{25}{24\pi} c^*
(2 \gamma \partial^3 \gamma - 3 \partial \gamma \partial^2 \gamma),
\end{equation}
where the dots now indicate the terms written in (\ref{s11}).
One finally checks that all the higher order equations both in $\hbar$
and in antifields are satisfied identically.

\section{Appendix}

\setcounter{equation}{0}

The transformations rules for the classical fields are given by
\begin{eqnarray}
\dd h &=& \bar{\partial} \epsilon - h \partial \epsilon
+ \epsilon \partial h + (\lambda \partial b - b \partial \lambda)
 \partial \chi^\alpha \partial \chi^\alpha \nonumber \\
&\ &+ \sqrt{\hbar} 2 \alpha_\alpha
[(\lambda \partial b - \partial \lambda b) \partial^2 \chi^\alpha
+ \partial \chi^*_\alpha
(\lambda \partial \gamma -  \gamma \partial \lambda)]
\nonumber \\
&\ &+\hbar \frac{25}{24}
(2 \lambda \partial^3 b - 3 \partial \lambda \partial^2 b
+3 \partial b \partial^2 \lambda - 2 b \partial^3 \lambda), \\
\dd b &=& \epsilon \partial b - 2 b \partial \epsilon
+ \bar{\partial} \lambda - h \partial \lambda + 2 \lambda \partial h, \\
\dd \chi^\alpha &=& \epsilon \partial \chi^\alpha
+ \lambda \partial \chi^\beta \partial \chi^\gamma d^{\alpha \beta \gamma}
-2 \hat{h}^* (\lambda \partial \gamma -  \gamma \partial \lambda)
\partial \chi^\alpha \nonumber \\
&\ &+ \sqrt{\hbar}[(-\alpha_\alpha \partial \epsilon +
(e_{\alpha \beta} - e_{\beta \alpha}) \lambda \partial^2 \chi^\beta
-e_{\beta \alpha} \partial \lambda \partial \chi^\beta) \nonumber \\
&\ & \hspace{0.7cm}
+2\partial[\hat{h}^*(\lambda \partial \gamma -  \gamma \partial \lambda)]
+\hbar f_\alpha \partial^2 \lambda, \\
\dd c &=& \epsilon \partial c  - c \partial \epsilon
+ (\lambda \partial \gamma -  \gamma \partial \lambda)
\partial \chi^\alpha \partial \chi^\alpha, \nonumber \\
&\ &+ \sqrt{\hbar}2 \alpha_\alpha
(\lambda \partial \gamma -  \gamma \partial \lambda) \nonumber \\
&\ &+\hbar \frac{25}{24}
(2 \gamma \partial^3 \lambda - 2 \lambda \partial^3 \gamma
- 3 \partial \gamma \partial^2 \lambda
+ 3 \partial \lambda \partial^2 \gamma), \\
\dd \gamma &=& \epsilon \partial \gamma - 2 \gamma \partial \epsilon
- c \partial \lambda + 2 \lambda \partial c.
\end{eqnarray}
The background fields transform the same way as the corresponding fields.

The transformations of the hatted ``antifields'' are the following
\begin{eqnarray}
\dd \hat{h}^* &=& \partial \hat{h}^* \epsilon
+ 2 \hat{h}^* \partial \epsilon +
2 \partial \hat{b}^* \lambda + 3 \hat{b}^* \partial \lambda, \label{hhat}\\
\dd \hat{b}^* &=&
\lambda \partial(\hat{h}^*\partial \chi^\alpha\partial \chi^\alpha)
+2 \hat{h}^* \partial \lambda \partial \chi^\alpha\partial \chi^\alpha
+ \partial \hat{b}^* \epsilon +3\hat{b}^* \partial \epsilon \nonumber \\
&\ &+2\sqrt{\hbar}\alpha_\alpha
[\lambda \partial (\hat{h}^* \partial^2 \chi^\alpha) +
2\hat{h}^* \partial \lambda \partial^2 \chi^\alpha] \nonumber \\
&\ &+\hbar \frac{25}{24}
(2 \partial^3 \hat{h}^* \lambda
+ 9 \partial^2 \hat{h}^* \partial \lambda
+ 15 \partial \hat{h}^* \partial^2 \lambda
+ 10  \hat{h}^* \partial^3 \lambda).
\end{eqnarray}

The ``genuine'' antifields transform as
\begin{eqnarray}
\dd \chi^*_\alpha &=& \partial \Big[ 2\hat{h}^*
(\lambda \partial B - B \partial \lambda )\partial \chi^\alpha
+2 c^* (\lambda \partial \gamma -  \gamma \partial \lambda)
 \partial \chi^\alpha \nonumber \\
&\ & \hspace{0.3cm}+\chi^*_\alpha \epsilon
+2 \lambda d^{\alpha \beta \gamma}\chi^*_\beta \partial \chi^\gamma
-2 \chi^*_\alpha \hat{h}^* (\lambda \partial \gamma -  \gamma \partial \lambda)
\nonumber \\
&\ & \hspace{0.3cm}
+\sqrt{\hbar}
[(e_{\alpha \beta} - e_{\beta \alpha})
\partial (\chi^*_\beta \lambda)
- e_{\beta \alpha} \chi^*_\beta \partial \lambda] \nonumber \\
&\ & \hspace{0.3cm}
- \alpha_\alpha \partial [\hat{h}^*(\lambda \partial b - b \partial \lambda)]
+2 \alpha_\alpha \partial
[c^*(\lambda \partial \gamma - \gamma \partial \lambda)]
 \Big], \label{chi} \\
\dd c^* &=& \partial c^* \epsilon + 2 c^* \partial\epsilon
+2 \partial \gamma^* \lambda + 3 \gamma^*\partial\lambda, \\
\dd \gamma^* &=&  \partial \gamma^* \epsilon + 3 \gamma^* \partial \epsilon
\nonumber \\
&\ &
+\lambda \partial(c^* \partial \chi^\alpha\partial \chi^\alpha)
+2 c^* \partial \lambda \partial \chi^\alpha \partial \chi^\alpha \nonumber \\
&\ &
-4 \chi^*_\alpha \hat{h}^* \partial \lambda \partial \chi^\alpha
- 2 \lambda \partial (\chi^*_\alpha \hat{h}^* \partial \chi^\alpha),
\nonumber \\
&\ &+\sqrt{\hbar} 2 \alpha_\alpha
[\lambda \partial (\hat{h}^* \partial \chi^\alpha)
+2 \partial \lambda \hat{h}^* \partial \chi^\alpha \nonumber \\
&\ &
+\lambda \partial(c^* \partial^2 \chi^\alpha)
+2 \partial \lambda c^* \partial^2 \chi^\alpha] \nonumber \\
&\ & +\hbar \frac{25}{24}
(-2 c^* \partial^3 \lambda + 3 \partial^2 c^* \partial \lambda
- 3 \partial c^* \partial^2 \lambda +2 \partial^3 c^* \lambda). \label{gamma}
\end{eqnarray}

Finally, we give the transformation rules for the BRST auxiliary fields
\begin{eqnarray}
\dd d_h &=& \partial d_h \epsilon + 2 d_h \partial \epsilon
+ 2 \partial d_b \lambda  + 3 d_h \partial \lambda, \\
\dd d_b &=& \lambda \partial (d_h \partial \chi^\alpha \partial \chi^\alpha)
+2 \partial \lambda d_h \partial \chi^\alpha \partial \chi^\alpha
+3 d_b \partial \epsilon + \partial d_b \epsilon \nonumber \\
&\ &+2 \sqrt{\hbar} \alpha_\alpha
[\lambda \partial (d_h \partial^2 \chi^\alpha)
+2 \partial \lambda d_h \partial^2 \chi^\alpha] \nonumber \\
&\ &+\hbar \frac{25}{24}
(2 \partial^3 d_h \lambda
+ 9 \partial^2 d_h \partial \lambda
+ 15 \partial d_h \partial^2 \lambda
+ 10  d_h \partial^3 \lambda).
\end{eqnarray}

\end{document}